# Service Mesh: Architectures, Applications, and Implementations

**Behrooz Farkiani** (A paper written under the guidance of Prof. Raj Jain)    Download 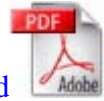

## Abstract

The scalability and flexibility of microservice architecture have led to major changes in cloud-native application architectures. However, the complexity of managing thousands of small services written in different languages and handling the exchange of data between them have caused significant management challenges. Service mesh is a promising solution that could mitigate these problems by introducing an overlay layer on top of the services. In this paper, we first study the architecture and components of service mesh architecture. Then, we review two important service mesh implementations and discuss how the service mesh could be helpful in other areas, including 5G.

## Keywords

Service Mesh, Cloud Native Application, Container, Kubernetes, Istio, Linkerd, 5G.

## Table of Contents

Service Mesh: Architectures, Applications, and Implementations

# 1. Introduction

A service mesh is a dedicated overlay layer on top of (micro) services that handles service-to-service communication. The main goal of the service mesh is the reliable delivery of requests through the topology of services. Although there is no official standard architecture of the service mesh concept and its components, researchers defined and proposed its components in both control and data planes. There are also two important implementations: Istio and Linkerd. We discuss these implementations and how service mesh could benefit other computing areas.

The structure of this paper is as follows. Section 2 briefly reviews the general components of service mesh architecture. We also review how the service mesh could be utilized in edge computing and Fifth-generation cellular technology (5G). Then, we discuss two important service mesh implementations named Istio and Linkerd and compare them with other implementations in Section 3. Finally, Section 4 concludes this paper.

# 2. Service Mesh

A cloud-native application might consist of several (micro)services that might be implemented in different programming languages, belong to different tenants, and have many service instances with a short lifetime to support traffic demands. It is the job of the service orchestrator component to manage this dynamic environment, manage and debug their interactions with each other and traffic flow, monitor their performance, and collect statistics related to the service [Li19][Redhat]. However, with a large number of services, efficiently performing these tasks becomes challenging.

Service mesh was introduced to mitigate the difficulty of performing the aforementioned tasks. In general, a service mesh implementation should provide the following features [Khatri20] [Li19]:

- Observability: The control should provide the observability of services running in the data plane. This could be done through distributed tracing [Cha21].
- Automatic scaling: The control plane services should automatically scale to handle the increased workload.
- Routing: The service mesh should manage the traffic routing rules between services running in the data plane and provide reliable delivery of messages. It should also use the gathered statistics to balance the load between different instances.
- Automatic service registration and discovery: In microservice applications, the number of service instances, their location, and their states are dynamically changing. The control plane should have the ability of automatic service discovery.
- Circuit breaking: in case of overloaded services, the circuit breaking feature should back off requests instead of allowing a wide system failure.

Service Mesh: Architectures, Applications, and Implementations

- Authentication and access control: A service mesh implementation should enforce service-to-service access policies.

In the rest of this section, we first review the background concepts. Then, we introduce a general service mesh architecture and continue to discuss the performance impact of implementing a service mesh. Finally, we discuss its business importance and the applications of service mesh in 5G and edge computing.

## 2.1. Background

This section explains the important concepts that will be used in the rest of this paper. Here, we review the evolution of software architecture, cloud-native applications, and service mesh definition.

### 2.1.1. The Evolution of Software Architecture

In recent years, we have witnessed a shift from monolithic applications architecture to service-oriented and microservice architecture. In monolithic architecture, all components of the application are tightly coupled together. As another approach, we could define and design services and break down the entire application into a set of services, each providing a business function; one of the main important features of service-oriented architecture is the loose coupling between service consumers and providers. These services could be developed, deployed, scaled, and administrated independently, and they have little or no knowledge of each other or any integration. Figure 1 and Figure 2 represent examples of monolithic and service-oriented architectures, respectively.

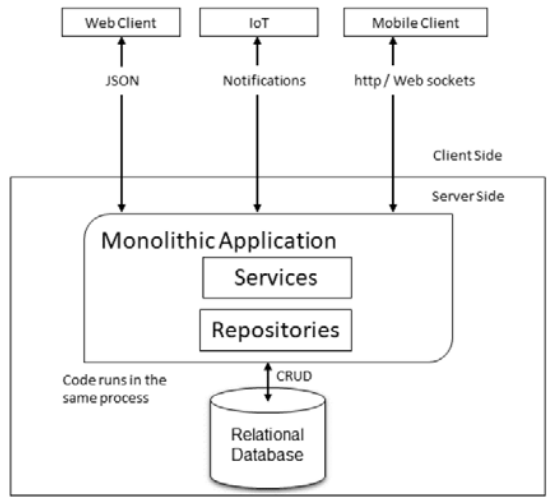

**Figure 1- Monolithic Applications [Khatri20]**

Service Mesh: Architectures, Applications, and Implementations

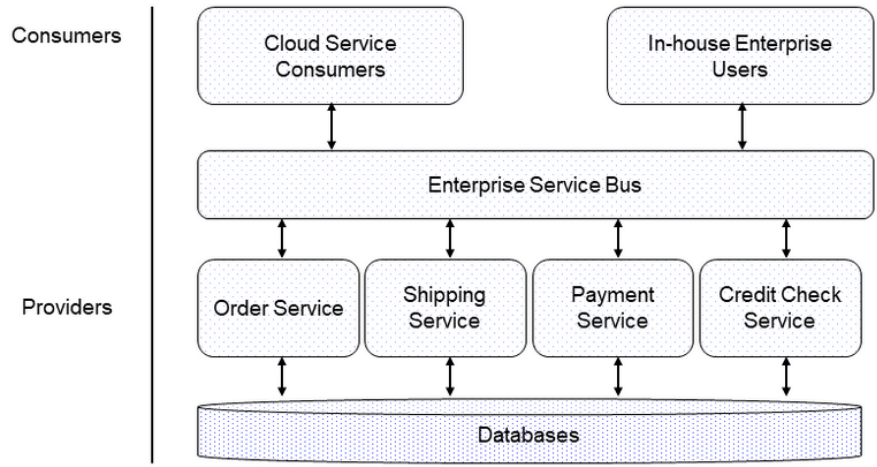

**Figure 2- Service Oriented Architecture [Khatri20]**

Microservice architecture is still a service-oriented architecture; it is made of reusable, loosely coupled (relatively small) components that work independently of each other. However, the main difference between these two architectures resides in their scopes: in service-oriented architecture, we focus on an enterprise scope, while in a microservice architecture, the focus is on the application level [IBM]. Figure 3 explains this difference.

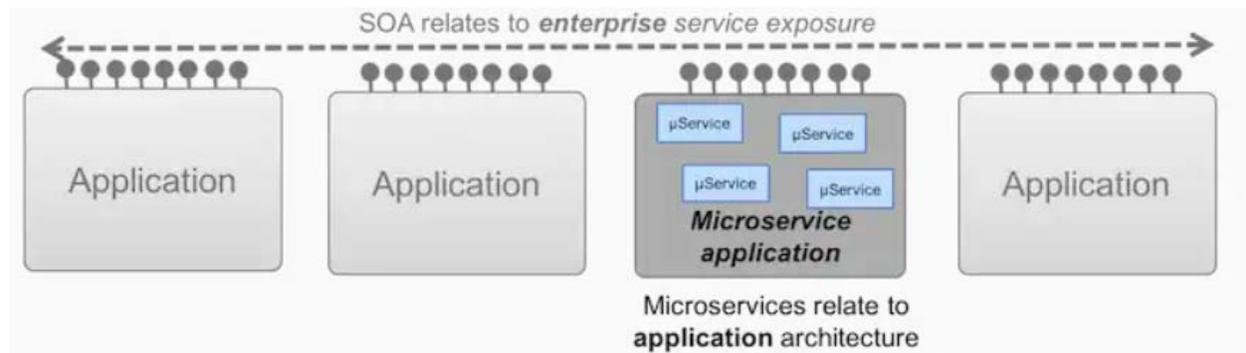

**Figure 3- Service oriented vs. microservice architecture [IBM]**

Other differences stem from this fact. Service-oriented architecture has a higher level of reuse and a lower level of data synchronization. Indeed, service-oriented architecture needs reuse and component sharing to achieve its scalability and efficiency goals. In addition, data is usually accessed and modified at its main source, which reduces the need for synchronization. On the other hand, reuse in microservice architecture leads to some level of dependency, which reduces agility and resilience. Therefore, we witness duplication of services. More importantly, each microservice has its own local copy of the data it needs. Two main advantages of microservice architecture are rapid development and a higher level of scalability in comparison to service-oriented architecture. An example of microservice architecture is shown in Figure 4.

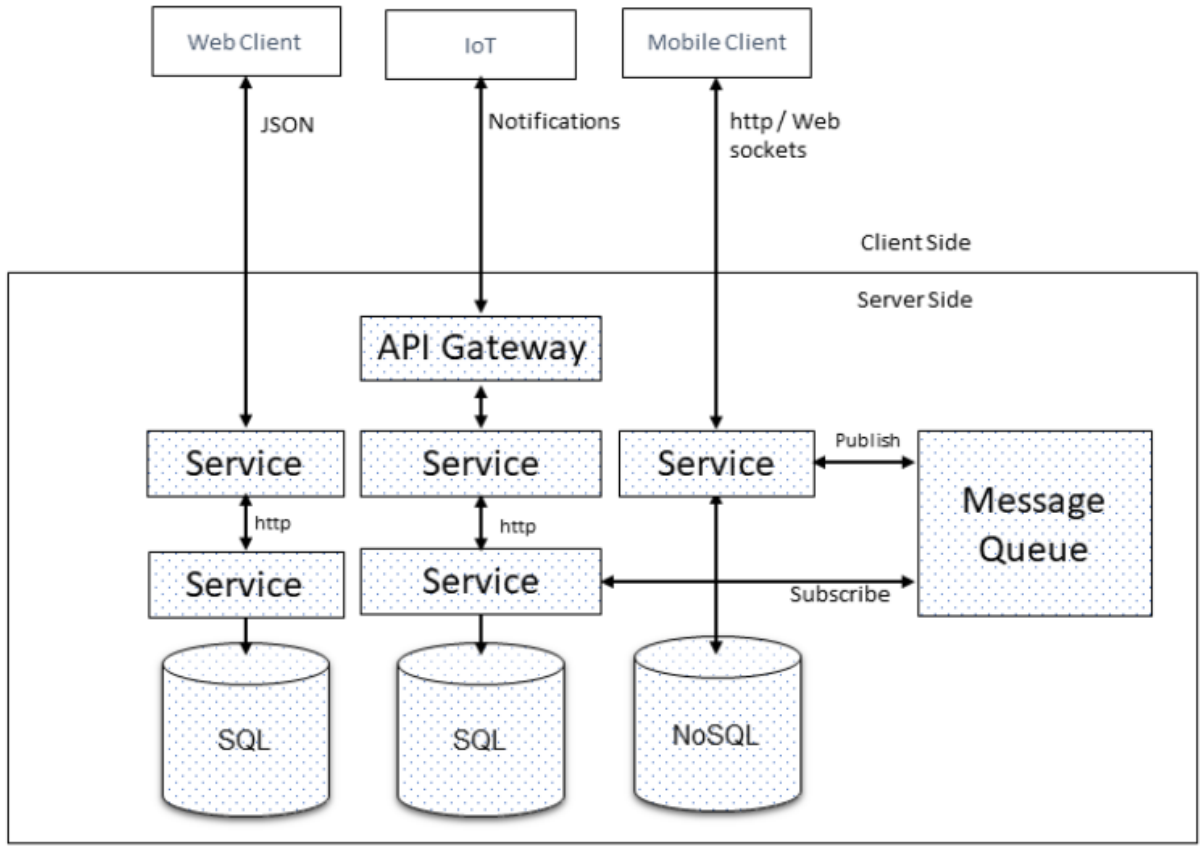

**Figure 4- Microservice architecture [Khatri20]**

### 2.1.2. Cloud-Native Applications

Another important concept is cloud-native applications, which has recently been used to describe container-based environments. Cloud-native applications usually refer to applications in which software development is a relatively rapid process because of the automated scalability and deployment process. An example of cloud-native applications is shown in Figure 5 [Khatri20].

Service Mesh: Architectures, Applications, and Implementations

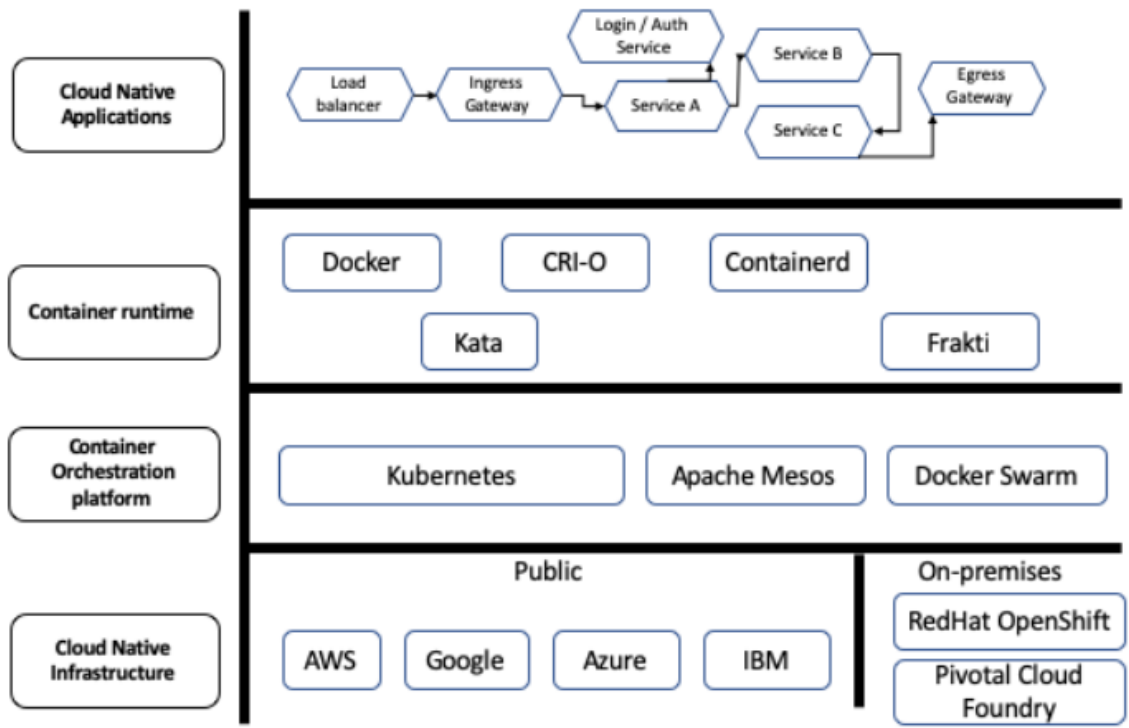

**Figure 5- Cloud Native Application [Khatri20]**

Cloud-native applications and microservice architecture usually benefit from containerized environments and run their services in containers. A container is a unit of software that packages up the code and all its dependencies. It enables us to move the application from one computing environment to another one and to run the application quickly and reliably [Docker]. We could create a container image that includes the program and its dependencies. A container runtime, like Docker, provides an environment to execute dockers on the host operating system. Containers are lighter than virtual machines and provide less isolation in comparison to them. In addition, in large production environments, we need a container orchestration platform to manage the life cycle of containers. Kubernetes [Kubernetes] is a very popular orchestration system that is widely used in the current implementation of service mesh architecture as the container orchestration platform.

**2.1.3. Envoy**

Envoy is a layer-7 proxy and communication bus designed for large modern service-oriented architectures. Envoy can shape, shift, split, route traffic, and collect telemetry for all service calls. Envoy proxy is transparent to applications and provides the following features [Khatri20] [Envoy]:

- **Out-of-process architecture**: This feature is also known as the **sidecar** proxy. It means that the Envoy proxy runs alongside the application and is language-agnostic.
- **Layer-3/Layer-4 filter**: At its core, Envoy is also an L3/L4 network proxy that provides a pluggable filter chain mechanism.

Service Mesh: Architectures, Applications, and Implementations

- **HTTP layer-7 filter**: There is an Hyper Text Transfer Protocol (HTTP) Layer-7 filter that supports buffering, rate limiting, and routing/forwarding.
- **HTTP/2 support**: Envoy supports both HTTP 1.1 and 2, and it can operate as a transparent HTTP/1.1 to HTTP/2 proxy in both directions.
- **HTTP layer-7 routing**: Envoy supports a routing subsystem that can route requests based on path, authority, content type, and runtime values.
- **gRPC support**: Google Remote Procedure Call (gRPC) is an RPC framework that uses HTTP/2. Envoy supports routing and load balancing for gRPC requests and responses.
- **Service discovery and dynamic configuration**: Envoy provides an optional dynamic configuration Application Programming Interface (API) for centralized management.
- **Health checking**: Envoy includes an active health-checking subsystem for upstream services. Envoy uses the collected information to determine healthy targets for load balancing.
- **Advanced load balancing**: Envoy is a self-contained proxy that could implement advanced load balancing techniques. Currently, it supports circuit breaking, automatic retries, global rate limiting, request shadowing, and outlier detection.
- **Front/edge proxy support**: Envoy's primary use is for service-to-service communication as a sidecar proxy. However, it can also act like an edge proxy because it supports HTTP/1.1, HTTP/2, HTTP/3, and HTTP layer-7 routing.
- **Best in class observability**: Envoy collects statistics for all subsystems.

## 2.2. Architecture

Service mesh is an application infrastructure layer on top of the microservice architecture that manages microservice-to-microservice communication. As a layer, it has both control and data planes. Service mesh is a concept, and there is no standard definition for it that specifies all its requirements and components. In this section, we present its components that are generally accepted and used in the research and software communities. Section 3 provides a detailed description of popular implementations.

Figure 6 illustrates different components of the service mesh. The following general components are suggested in each of the data and control planes.

- **Data plane**: The main data plane component is the "sidecar" proxy. These proxies are deployed independently beside every service component and are invisible to the services they are attached to. Through these proxies, the control plane could manage service mesh.
- **Control plane**: The control plane manages the configurations, policies, and management services. The control plane also provides secure communication between microservices through authentication and authorization services.

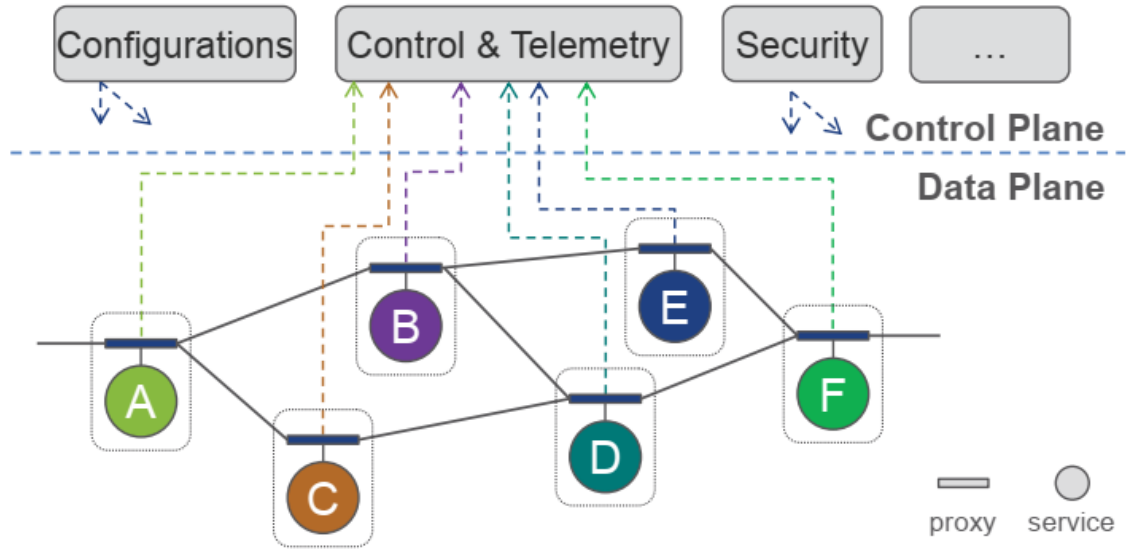

**Figure 6- Service Mesh Architecture [Li19]**

## 2.3. Performance Impact

In service mesh, traffic passes through additional sidecar proxies. This will result in additional end-to-end delay and reduces the performance. Authors of [Zhu22] designed a decomposition approach and a tool named MeshInsight to measure service mesh overhead. They showed implementing service mesh could result in 185% higher latency and 92% more virtual cores.

Figure 7 shows the data path for both inbound and outbound traffic. We could see there are three separate connections: two between sidecar proxies and their microservices and one between sidecar proxies. As a source of overhead, the message buffer should also be copied into the proxy buffer and vice versa. In addition, there are additional system calls, and the sidecar may process the message layers to do some actions. The authors also mentioned that protocol parsing is a major source of overhead for HTTP and gRPC proxies.

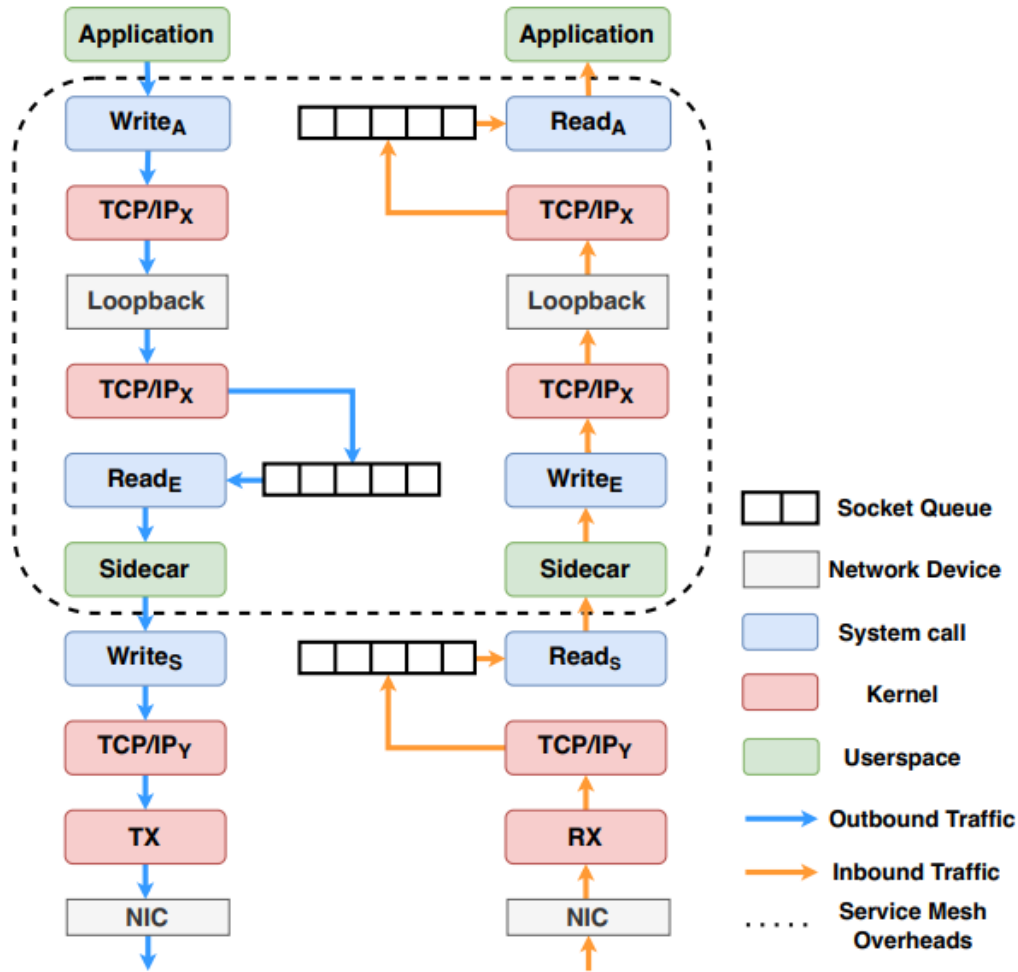

**Figure 7- Service mesh proxy data path [Zhu22]**

The authors of [Ganguli21] measured the performance impact of deploying service mesh in edge environments. They deployed Kubernetes in a virtual machine environment instead of a containerized environment. The authors showed that using Istio could reduce HTTP throughput between virtual machines by up to 70% and could double the tail latency.

## 2.4. Business Importance

Gartner [Gartner] categorized service mesh as an adolescent technology with 1% to 5% market penetration. It also categorized service mesh in the "Trough of Disillusionment" phase of the hype cycle. This means that the producers of the service mesh need to shake out or fail; they need to improve their products to satisfy early adopters.

## 2.5. Applications

This section reviews the application of service mesh in other areas. For example, [XIE20] proposed using Kubernetes and Istio for an on-demand image classification application to support load balancing and scheduling. Here, we discuss how we could employ service mesh to improve 5G network efficiency and how it could be used in resource-limited edge computing environments.

**2.5.1. 5G**

The authors of [Dab20] investigated the problem of steering traffic between microservice-based network functions in 5G architecture. One way to achieve the requirements of 5G networks while reducing the total operational costs is to use cloud-native applications. However, steering traffic between network functions is challenging, and service mesh could help us to tackle this issue. The authors proposed a cloud-native service function chaining framework based on Kubernetes and Network Service Mesh [NSM]. Then, the authors formulated the network-aware load-balancing optimization problem and proposed an algorithm to solve it.

[Wojciechowski21] proposed a scheduler for 5G networks by extending the Kubernetes scheduler and utilizing information gathered by the Istio service mesh. The authors aimed to improve service placement to reduce the latency. Their scheduler uses two metrics that are gathered by the service mesh: the number of bytes that are transferred in requests and responses. The scheduler uses these metrics to calculate the average flow between applications. Then, it could detect the nodes that have the highest flow and collocate them.

**2.5.2. Edge Computing**

[Furusawa22] proposed a service mesh controller that balances the load between edge servers. Usually, the service mesh is used in cloud environments. However, we could benefit from deploying them in edge environments because of the limited computing resources of edge servers. Consider a set of edge servers hosting applications that serve cars. In the case of car accidents and traffic congestion, the requests to the servers that are located in the related geographical area increase, and these servers become overloaded. In such cases, cooperative load balancing could be beneficial to avoid edge server overloading. In cooperative load balancing, some requests are redirected to other nearby servers. However, the current Kubernetes container execution implementation lacks the feature of using geographical data that is essential to implement cooperative load balancing. Therefore, the authors utilized Istio and proposed a weight calculation algorithm that is used to forward traffic to other nearby edge servers. As a systematic review of the challenges of implementing service mesh in edge environments, readers should consult [Duque22].

**2.6. Summary**

We reviewed the shift in software architecture from the monolithic architecture to microservices. Then, we discussed the main features of the service mesh, its performance impact, and its business importance. While employing service mesh could ease the management of microservices, it also could lead to significant performance degradation. As it is noted by [Sedghpour22], employing eBPF for root cause analysis, high-performance monitoring, and management could significantly improve the performance of service mesh implementations. We also reviewed the applications of service mesh in 5G and edge computing.

Service Mesh: Architectures, Applications, and Implementations

# 3. Implementations

Two important and widely used implementations of service mesh are Istio and Linkerd. In this section, we discuss these implementations and their components. Then, we provide a table to compare different service mesh implementations.

## 3.1. Istio

Istio service mesh, started in May 2017, is one of the fastest-growing open-source service mesh projects. Istio extends Kubernetes and utilizes Envoy proxies to provide traffic management, telemetry, and security [Istio].

Istio has a centralized control plane and supports integration with virtual machines and service discovery through other third-party service catalogs. Istio uses Envoy as its sidecar proxy and extends the Kubernetes API server for configuration management and access control. It also uses Kubernetes' built-in datastore, called etcd, to store its state and configuration. A high-level view of Istio architecture is shown in Figure 8 [Istio] [Khatri20].

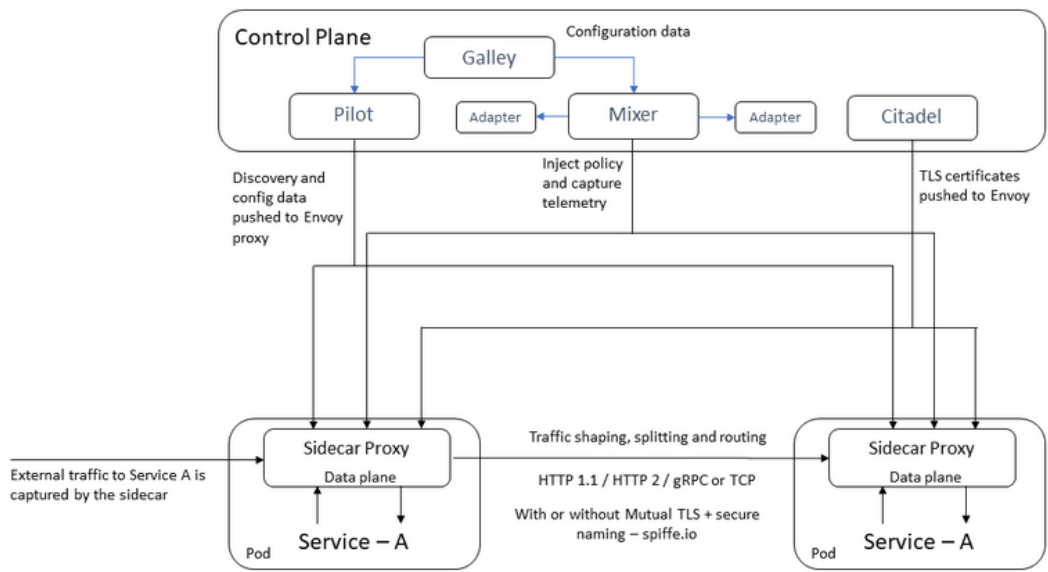

**Figure 8- Istio Architecture [Zhu22]**

The Istio control plane has four main components:

- Gallet
- Pilot

Service Mesh: Architectures, Applications, and Implementations

- Mixer
- Citadel

### 3.1.1. Galley

Galley gathers and validates user configuration for the other parts of the system. Galley provides configuration management services to different Istio components.

### 3.1.2. Pilot

Pilot is the traffic management component of Istio. It pushes communication-based policies to sidecar proxies at runtime to enforce traffic management configurations. Pilot maintains an abstract model of all of the services in the mesh that have been discovered through Kubernetes or Gallery. The platform-specific adapters, such as Kubernetes, are used to populate the abstract model with the service registry and resource information. Kubernetes stores the service discovery metadata in the etcd database when we create Kubernetes services. Figure 9 illustrates the components of Pilot.

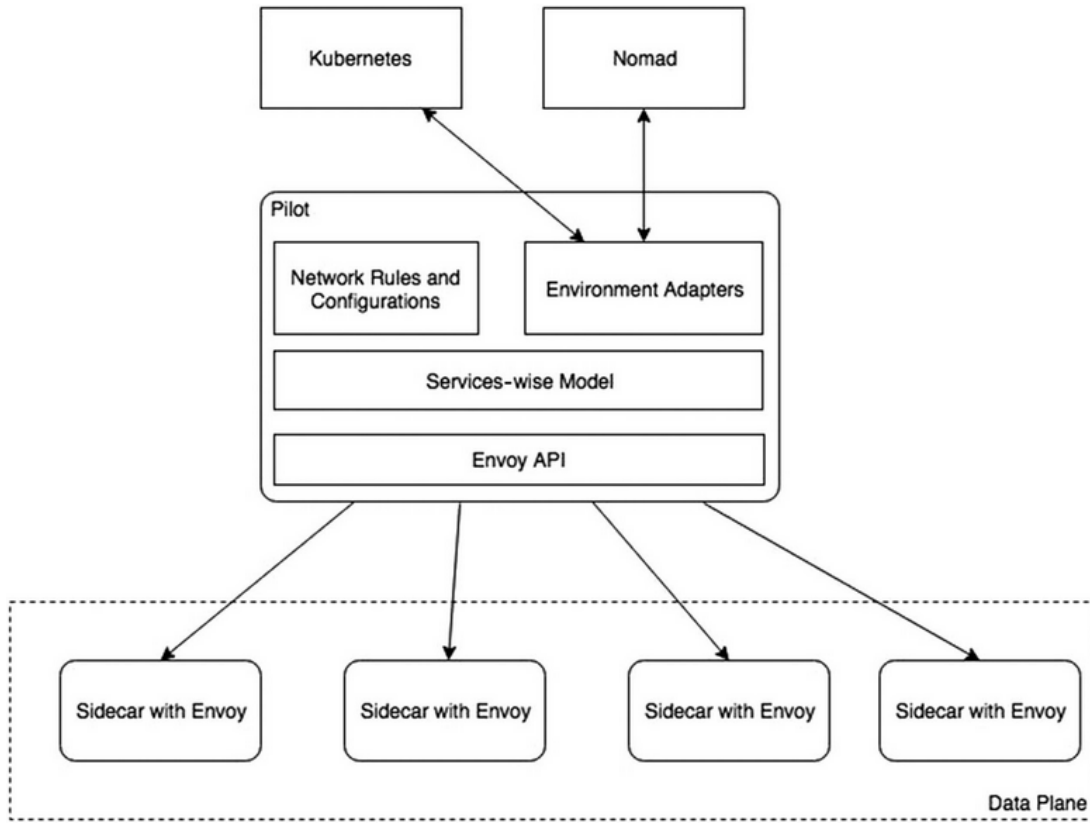

**Figure 9- The components of Pilot [Sharma19]**

Service Mesh: Architectures, Applications, and Implementations

### 3.1.3. Mixer

Istio Mixer component is a general-purpose policy and telemetry hub. Mixer enables access control and manages authorization and auditing, telemetry capturing, and quota enforcing. The components of the Mixer are shown in Figure 10.

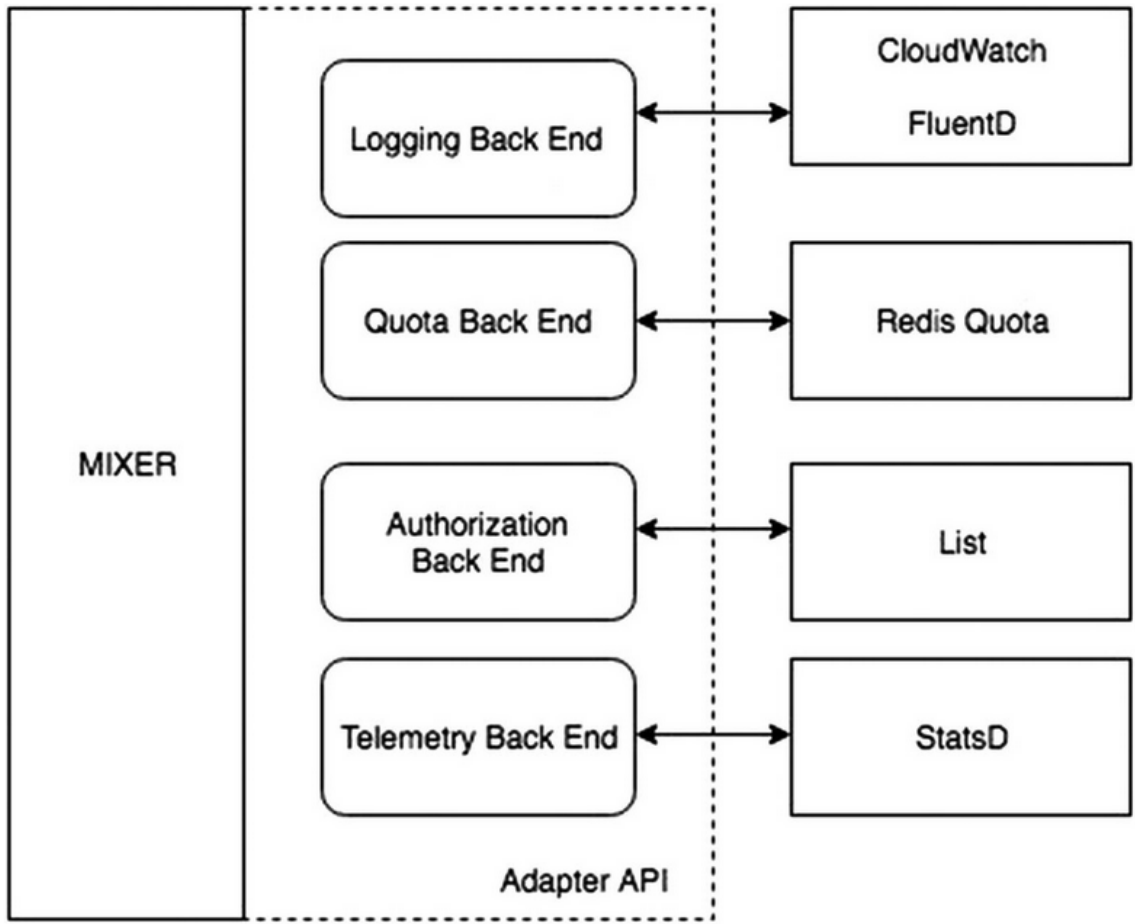

**Figure 10- Mixer Components [Sharma19]**

### 3.1.4. Citadel

Istio Citadel component enables service-to-service and end-user authentication and identity management. The Istio security model is implemented through the following control plane components:

- **Citadel** manages keys and certificates.
- **Pilot** distributes the authentication policies.
- **Mixer** provides authorization and auditing policies.

Service Mesh: Architectures, Applications, and Implementations

## 3.2. Linkerd

Linkerd [Linkerd] advertises that it is the fastest and lightest implementation of the service mesh on top of Kubernetes. Linkerd does not use Envoy. Instead, it uses its own lightweight layer-7 micro proxy named Linkerd2-proxy as the sidecar proxy. The proxy sits next to every microservice, wraps the network call, and collects the metrics. Linkerd encrypts all service-to-service communication through Transport Layer Security (TLS), and all the traffic on the wire is also encrypted. Linkerd provides load balancing, TLS, request routing, and service scalability. A high-level diagram of Linkerd architecture is shown in Figure 11 [Khatri20].

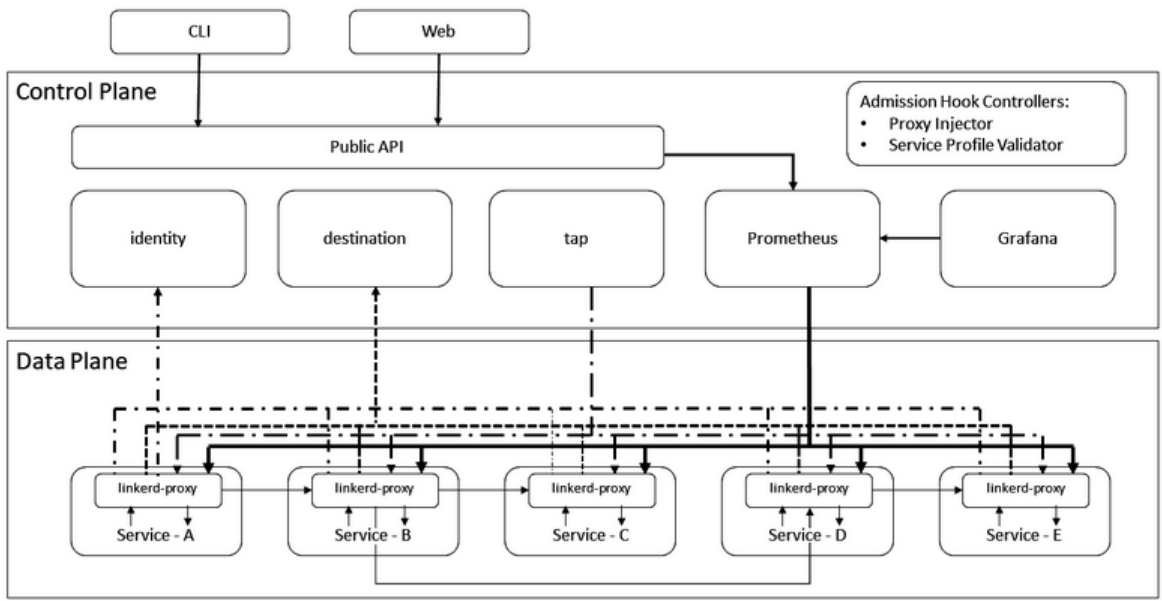

**Figure 11- Linkerd Architecture [Khatri20]**

The primary functions of the Linkerd control plane are telemetry data aggregation, service API calls, and enabling data access between the control plane and service proxies. The control plane has the following components:

- **Identity**: The identity component is a TLS certificate authority that manages keys for proxy-to-proxy connections to implement Mutual TLS (mTLS).
- **Destination**: It is used to fetch service discovery information, policy information, and service profile information.
- **Prometheus**: Stores metrics, telemetry, and monitoring data that have been captured by Linkerd proxies and metrics that other Linkerd components have generated.
- **Grafana**: It integrates with Prometheus to visualize metrics that Prometheus has captured.
- **Tap**: Allows introspection of live traffic in real-time. Access to it is controlled using role-based access control.

Service Mesh: Architectures, Applications, and Implementations

## 3.3. Comparison of Implementations

We use the table presented in [serviceComp] to compare the features of Istio [Istio], Linkerd [Linkerd], Consul [Consul], Kuma [Kuma], and Open Service Mesh [OSM]. Table 1 describes these differences.

**Table 1- Comparison of different service mesh implementations [serviceComp]**

| Feature \ Implementation | Istio | Linkerd | Consul | Kuma | Open Service Mesh |
|---|---|---|---|---|---|
| **License** | Apache License 2.0 | Apache License 2.0 | Mozilla License | Apache License 2.0 | Apache License 2.0 |
| **Service Proxy** | Envoy, proxyless for gRPC (experimental) | Linkerd2-proxy | defaults to Envoy, exchangeable | Envoy | Envoy |
| **TCP** | Yes | Yes | Yes | Yes | Yes |
| **HTTP/1.1+** | Yes | Yes | Yes | Yes | Yes |
| **HTTP/2** | Yes | Yes | Yes | Yes | Yes |
| **gRPC** | Yes | Yes | Yes | Yes | Yes |
| **Automatic Sidecar Injection** | Yes | Yes | Yes | Yes | Yes |
| **Platform** | Kubernetes | Kubernetes | Kubernetes, Nomad, VMs, ECS, Lambda | Kubernetes, VMs, ECS | Kubernetes |
| **Extension of the Mesh by containers/VMs outside the cluster** | Yes | No | Yes | Yes | No |
| **Control and observe multiple clusters** | Yes | Yes | Yes | Yes | planned |
| **Traffic Access Control** | Yes | No | Yes | No | Yes |

**Table 1- Comparison of different service mesh implementations [serviceComp]**

| Feature \ Implementation | Istio | Linkerd | Consul | Kuma | Open Service Mesh |
|---|---|---|---|---|---|
| **Traffic Split** | Yes | Yes | No | No | Yes |
| **Traffic Metrics** | Yes | Yes | No | No | Yes |
| **Service Log Collection** | No | No | No | No | Yes, using Fluent Bit |
| **Access Log Generation** | Yes | No (tap feature instead) | Yes | Yes | No |
| **Per-Route Metrics** | experimental | Yes | depending on the proxy used | No | No |
| **Load Balancing** | Yes (Round Robin, Random, Weighted, Least Request) | Yes (exponentially weighted moving average) | Yes (Round Robin, Random, Weighted, Least Request, Ring Hash, Maglev) | Yes (Round Robin, Least Request, Ring Hash, Random, Maglev) | Yes |
| **Percentage-based Traffic Splits** | Yes | Yes | Yes | Yes | Yes |
| **Header- and Path-based Traffic Splits** | Yes | planned | Yes | Yes | Header-based |
| **Circuit Breaking** | Yes | No, planned for 2.12.0 | Yes | Yes | Yes |
| **mTLS** | Yes | Yes, on by default | Yes | Yes | Yes |
| **mTLS Enforcement** | Yes | Yes | Yes | Yes | Yes, via https://linkerd.io/2.11/features/server-policy/ |
| **mTLS Permissive Mode** | Yes | Yes | No | Yes | Yes |

**Table 1- Comparison of different service mesh implementations [serviceComp]**

| Feature \ Implementation | Istio | Linkerd | Consul | Kuma | Open Service Mesh |
|---|---|---|---|---|---|
| mTLS by default | Yes, permissive mode | Yes, permissive mode | Yes | No | Yes |
| Service-to-Service Authorization Rules | Yes | Yes | Yes | Yes | Yes |

### 3.4. Summary

In this section, we reviewed the popular service mesh implementations and their components. Then, we provided a table that compares the different features of the five most important implementations. It is clear that all of the reviewed service meshes rely on Kubernetes as the orchestration framework. Therefore, any future implementation should also consider Kubernetes as one of the candidates for the orchestration system.

## 4. Summary

In this paper, we reviewed the service mesh concept, its features, and its popular implementation. We showed that implementing service mesh could ease service management and policy enforcement and improve service observability. It also enables us to extend orchestration framework capabilities without directly modifying its core code. However, it also results in performance degradation that needs to be addressed. We also discussed the benefits of using service mesh in 5G and edge environments.

It seems that current implementations of service mesh have reached a maturity level in terms of features. Therefore, future research should focus on how we could improve the performance of the implementations without limiting its functionality.

## 5. List of Acronyms

**Table 2- List of acronyms**

| | |
|---|---|
| 5G | Fifth-generation technology standard for broadband cellular networks |
| HTTP | Hypertext Transfer Protocol |
| TLS | Transport Layer Security |

Service Mesh: Architectures, Applications, and Implementations

**Table 2- List of acronyms**

| 5G | Fifth-generation technology standard for broadband cellular networks |
|---|---|
| mTLS | Mutual TLS |
| gRPC | Google Remote Procedure Call |
| API | Application Programming Interface |

Service Mesh: Architectures, Applications, and Implementations

---

Last modified on December 7, 2022
This and other papers on recent advances in Wireless and Mobile Networking are available online at http://www.cse.wustl.edu/~jain/cse574-22/index.html
Back to Raj Jain's Home Page